\documentclass[prl,twocolumn,showpacs,preprintnumbers,amsmath,amssymb]{revtex4}
\usepackage{graphicx}
\usepackage{epsfig}
\def\figref#1{Fig.~\ref{#1}}
\def\rl{{\bf r}_{\rm L}}  
\def\vl{{\bf v}_{\rm L}}  
\def\vs{{\bf v}_{\rm s}}  

\begin{document}
\title{Inertial Josephson Relation for FIR Frequencies}

\author{Pei-Jen Lin}
\email{fareh.lin@gmail.com}
\affiliation{NCTS, National Tsing Hua University, Hsinchu 300, Taiwan}
\author{P. Lipavsk\'y}
\email{lipavsky@fzu.cz}
\affiliation{Faculty of Mathematics and Physics, Charles University,
Ke Karlovu 3, 12116 Prague 2, Czech Republic}
\affiliation{Institute of Physics, Academy of Sciences,
Cukrovarnick\'a 10, 16253 Prague 6, Czech Republic}
\author{Peter Matlock}
\email{pwm@induulge.net}
\affiliation{Research Department, Universal Analytics Inc., Airdrie, AB, Canada}

\pacs{PACS number}

\begin{abstract}
Consideration of the balance of forces on superconducting condensate
at low frequencies leads to the well-known Josephson Relation.  Using
the Ginzburg-Landau expression for the current, an expression relating
the electric field to the vortex velocity via the magnetic field is
obtained. This result is the Josephson Relation, supplemented by a
term accounting for the inertia of charge carriers. This Inertial
Josephson Relation may be used at all frequencies and may be viewed as
the Josephson Relation extended to the case of sub-gap high-frequency
response. When applied to vortex dynamics it yields the same
conductivity as solution of the Ginzburg-Landau theory.
\end{abstract}

\maketitle

Vortices in the superconducting condensate are stable objects, which
interact to form vortex matter \cite{BFGLV94,RL09}. The response of
vortex matter may be studied at various frequencies.  One potential
practical application of vortex matter is as a memory device; one
might imagine that data could be encoded in the positioning of
vortices.  To attain this degree of control over vortex matter, it
must be clear theoretically how forces and fields interact with
vortices at the desirable high frequencies of such a putative memory
device.

Low-frequency vortex motion can be observed by magnetic scanning
\cite{TAK99}, but high frequencies are invisible via this
method. Fortunately, in far infra-red (FIR) spectroscopy it has been
possible to make high-frequency observations of the electric current
to which the vortex velocity responds \cite{GR66}.  Although these
data have then been interpreted using the Josephson Relation
\cite{J65}, one may ask whether this is legitimate at high
frequencies.

\begin{figure}[ht]
  \centerline{
  \psfig{figure=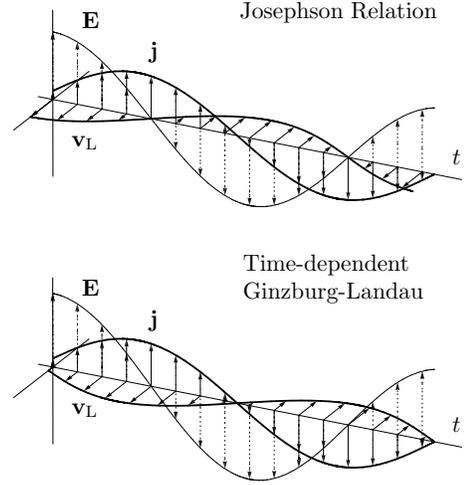,width=6cm,angle=0}}
\caption{Phases of electric field $\bf E$, current $\bf j$ and vortex velocity $\vl$ as functions of time $t$}
\label{fig_phases}
\end{figure}

\begin{figure}[ht] 
  \centerline{
  \psfig{figure=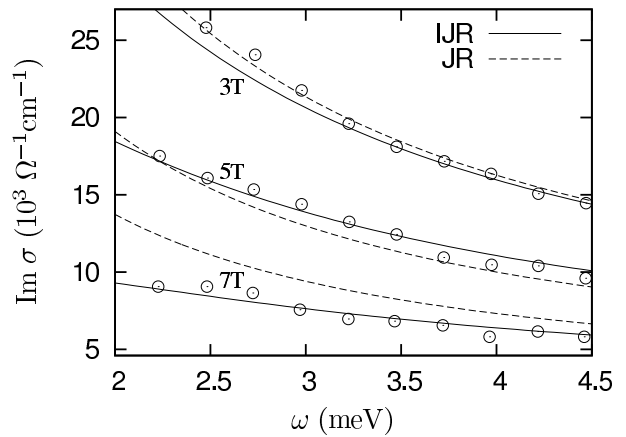,width=8cm}
  }
\vskip 2pt
\caption{Imaginary part of the conductivity for various magnetic field strengths: 
The points are experimental data of Ikebe {\em et al} \cite{ISIFK09}.
The Gittleman and Rosenblum model, using the Josephson Relation, 
is plotted with pinning $\nu=5.9$ ${\rm N}/{\rm cm}^2$.
The result from TDGL theory is coincident with the use of the Inertial 
Josephson Relation, without pinning.
}
\label{fig_dataplot}
\end{figure}

An alternative way to interpret the FIR data is based on the
Time-Dependent Ginzburg-Landau (TDGL) theory \cite{MRSS09,LM10}.
These two approaches lead to contradictory results; there is a
difference in phases which is illustrated schematically in
\figref{fig_phases}.  Actual experimental data is shown in
\figref{fig_dataplot}. Both approaches can reproduce the
experimentally established frequency-dependent phase difference
between the electric field and the current, but they yield different
vortex velocities. According to the Josephson Relation
\begin{equation}
\label{Josephson}
{\bf E}=-\frac{1}{c} \vl\times{\bf B}
\end{equation}
the vortex velocity $\vl$ is in phase with the electric field ${\bf  E}$. 
In contrary, in the solution \cite{LM10} of the TDGL equation
derived near the transition line, the vortex velocity is in phase with
the current.

These different phases of the vortex velocity lead to different
interpretations of the experimental data. The motion of vortices is
given by the balance of forces (per unit length)
\begin{equation}
\label{BoF}
\frac{\Phi_0}{cB}{\bf j}\times{\bf B} - \nu\rl=\eta\vl
,\end{equation}
where the left-hand side includes the Lorentz force due to the mean
current ${\bf j}$ and the pinning force proportional to the vortex
displacement $\rl$, and the right hand side is a friction
force. Identification of these forces corresponds to the Gittleman and
Rosenblum (GR) model \cite{GR66} which is sufficient for our
discussion.  Using the Josephson Relation \eqref{Josephson} one must
interpret the experimentally observed phase difference between the
field and current as a manifestation of the pinning force,
$\nu\not=0$.  Interpreting this same case using the TDGL theory, one
would not require the addition of a pinning potential.

The goal of the present paper is to show that the descrepancy
discussed above is actually a consequence of the inapplicability of
the Josephson Relation at high frequencies. Using the current from the
GL theory, we will show how one can derive the Inertial Josephson
Relation, which is similar to the Josephson Relation except for a term
important only at high frequencies, which is proportional to the
acceleration of charge carriers.  The Inertial Josephson Relation has
in fact been known in the literature for some time \cite{HMS92,AKK65},
but has been derived by much less direct means than that which we
propose here \cite{HMS92}.  In \cite{HMS92,AKK65} it is motivated by
the application of hydrodynamics to the superfluid state \cite{BK61}.

Although our arguments are in many respects more widely applicable,
for the sake of simplicity we restrict our attention to a
two-dimensional sample with perpendicular magnetic field ${\bf B}$
creating the triangular Abrikosov vortex lattice. We will derive the
relation for an ideal sample with no pinning centers. Electrons are
driven by a FIR light which we represent by a homogeneous electric
field parallel to the sample ${\bf E}= {\bf\tilde E}\cos\omega t$. The
sample is of infinitesimal thickness, therefore currents in the sample
have negligible feedback effect on the electromagnetic fields acting
on the sample. We neglect contributions important for the Faraday
rotation or the Hall voltage. Finally, we work in the vector gauge
with zero scalar potential, allowing us to avoid a distracting
discussion of the electrostatic and electrochemical potentials.

In principle one must solve for the order parameter $\psi$ from the
TDGL equation
\begin{equation}
\label{tdgl}
\frac{1}{2m^*}\bigg(
                   -i\hbar\nabla-\frac{e^*}{c}{\bf A}
              \bigg)^2 \psi 
+\alpha\psi+\beta|\psi|^2\psi=-
\Gamma \frac{\partial}{\partial t}\psi
\end{equation}
and then evaluate the current
\begin{equation}
\label{current}
{\bf J}=-\frac{e^{*2}}{ m^*c}{\bf A}|\psi|^2+
\frac{i\hbar e^*}{2m^*}(  \psi\nabla\bar\psi - \bar\psi\nabla\psi )
.\end{equation}
In the limiting case of linear response these two equations can be
mapped to equations \eqref{BoF} and \eqref{Josephson},
respectively. As shown in \cite{Kop01}, the TDGL equation \eqref{tdgl}
implies the balance of forces on vortices \eqref{BoF} with $\nu=0$.  A
second independent equation must be derived from the current formula
\eqref{current}. The electric field appears in the time derivative of
the current, since ${\bf E}=-(1/c)(\partial{\bf A}/\partial t)$.

In the linear approximation the vortex lattice is undistorted, and the
order parameter reflects this as a function of time.  We denote by
$\psi_0({\bf r})$ the order parameter in the equilibrium and by
$\psi({\bf r},t)$ its time-dependent value driven by the light.  The
coherent movement of the vortex lattice implies 
$\psi({\bf r},t)= {\rm  e}^{-i({\bf C}_{\rm L}\cdot{\bf r})}\psi_0({\bf r}-\rl)$, 
where $\rl$ is a displacement of the vortex lattice at time $t$ from its
equilibrium position. The phase factor is given by the relative
position of the displaced vortex lattice and the center of the vector
gauge, ${\bf C}_{\rm L}=(\pi/\Phi_0) {\bf B}\times \rl$.  Here
$\Phi_0=2\pi\hbar c/e^*$ is the elementary flux.  When calculating
$\partial{\bf J} / {\partial t}$, the time derivative can thus be
written in terms of space derivatives; 
$\partial\psi/\partial t=i(2\pi/\Phi_0)(\vl \cdot{\bf A})\psi-(\vl\cdot\nabla)\psi$, 
where we have used $\vl=(\partial\rl/ \partial t)$ and assumed $\vl$ parallel
to $\rl$.  We find
\begin{eqnarray}
\label{dJdt}
\frac{\partial{\bf J}}{\partial t}&=&
\frac{e^{*2}}{m^*}{\bf E}|\psi|^2 
+\frac{e^{*2}}{m^*c}{\bf A}{\vl}\cdot\nabla |\psi|^2 \\
&+&\frac{\hbar e^*}{2m^*}
   \bigg(
      i{\vl}\cdot\nabla (\psi^*\nabla\psi - \psi\nabla\psi^*)
      +\frac{4\pi}{\Phi_0} |\psi|^2 \nabla({\vl}\cdot{\bf A})
   \bigg) \nonumber
.\end{eqnarray}

The transport current ${\bf j}$ is the mean value
obtained by averaging over the elementary cell of the lattice.
We thus average the time derivative of the current 
\eqref{dJdt} and express the mean current via the Cooper pair velocity, 
${\bf j}=e^*n\vs$. Here 
$n=\bigl\langle|\psi|^2\bigr\rangle=(B/\Phi_0)\int_{\rm cell}d{\bf r}|\psi|^2$ 
is the mean density of Cooper pairs.
\begin{eqnarray}
\label{dvsdt}
\frac{m^*}{e^*}\frac{\partial\vs}{\partial t}&=&
{\bf E}
+{\frac{i\hbar}{2e^*n}} \big\langle {\vl}\cdot\nabla (\psi^*\nabla\psi-\psi\nabla\psi^* )\big\rangle\nonumber\\
&+&\frac1{cn} \big\langle {\bf A} {\vl}\cdot\nabla |\psi|^2 + |\psi|^2\nabla({\vl}\cdot{\bf A})\big\rangle
.\end{eqnarray}
Simplifying,
\begin{eqnarray}
\label{dvsdt2}
\frac{m^*}{e^*}\frac{\partial\vs}{\partial t}&=&
{\bf E}
-\frac{m^*}{{e^*}^2n}\big\langle {\vl}\cdot \nabla {\bf J} \big\rangle \nonumber\\
&+&\frac1{cn} \bigg\langle |\psi|^2 \big(\nabla ({\vl}\cdot{\bf A}) - {\vl}\cdot\nabla {\bf A} \big)\bigg\rangle
.\end{eqnarray}
The total derivative of the current is zero under integration, 
since the current itself is periodic on the lattice.
Using the identity 
${\bf b} {\bf a} \cdot {\bf c} - {\bf c} {\bf a} \cdot {\bf b} = {\bf a}\times ( {\bf b} \times {\bf c} )$
we thus obtain the relation
\begin{equation}
\label{inerJos}
\frac{m^*}{ e^*}\frac{\partial\vs}{\partial t}= {\bf E}+\frac{1}{c} \vl\times{\bf B}
\end{equation}
which is the Inertial Josephson Relation; we view it as extending 
the validity of the Josephson Relation \eqref{Josephson} into the FIR region.

In Kopnin \cite{Kop01} is given a derivation of the Josephson Relation
\eqref{Josephson} under certain rather broad assumptions.  It is
useful at this stage to make contact with this well-known result, and
show how this is commensurate with the Inertial Josephson Relation
\eqref{inerJos}.  Differences between \cite{Kop01} and the present
paper include Kopnin's use of gauge-invariant quantities, lack of
assumption of homogeneous electric field, and use of the phase $\chi$
of the order parameter $\psi=|\psi|e^{i\chi}$ explicitly in his
equations.  None of these differences is significant in the present
context. We choose to make a certain choice of gauge, but the final
result is one between physical quantities.  The simplicity of our
calculation is increased slightly due to our assumption of a
homogeneous electric field, since we may take ${\bf E}$ outside
$\langle\cdots\rangle$ when calculating the averaged current 
${\bf j}$.  Kopnin uses the $\chi$ field, which is subject to nonanalytic
behaviour due to a coordinate singularity when $|\psi|=0$, but this is
done carefully and no descrepancy is to be found; we use the complex
order parameter $\psi$ itself, which is not singular.

The essential difference is as follows. Kopnin assumes the vortex
configuration to move coherently at constant velocity, so that in the
notation of \cite{Kop01} (except that there $\Delta$ is used in place
of $\psi$), the vector potential is taken to be
${\bf A}({\bf r},t)={\bf A}_{\textup{static}}({\bf r}-{\vl} t) + {\bf A}_1$
and the order parameter
$\psi({\bf r},t)=\psi_{\textup{static}}({\bf r}-{\vl} t) + \psi_1$
where ${\bf A}_1$ and $\psi_1$ are taken to be small corrections.
No compensating gauge transformation is needed here as the calculation
is not performed in a particular gauge.  In fact, in the
time-dependent case these would need to be modified to read
\begin{equation}
{\bf A}({\bf r},t)={\bf A}_{\textup{static}}({\bf r}-{\rl}(t)) + {\bf A}_1
\end{equation}
and 
\begin{equation}
\psi({\bf r},t)=\psi_{\textup{static}}({\bf r}-{\rl}(t)) + \psi_1
\end{equation}
In addition, the longitudinal part of ${\bf A}_1$ may not be neglected, and we 
must insert the current
\begin{equation}
{\bf j} = - \frac{1}{c} \langle {\bf A}_1 |\psi|^2 \rangle \frac{{e^*}^2}{m^*}
.\end{equation}
With these modifications, relaxing the assumption of constant-velocity
motion of the vortices, the derivation of Kopnin reproduces precisely 
the Inertial Josephson Relation \eqref{inerJos}.


Let us illustrate in experimental context how the Inertial Josephson
Relation \eqref{inerJos} leads to conclusions different from those
based on the Josephson relation \eqref{Josephson}.  In
\figref{fig_dataplot} is shown the imaginary part of conductivity
observed with Fast FIR Transmission Spectroscopy by Ikebe {\em et al}
\cite{ISIFK09}.  With
${\bf j} = {\rm Re} ({\bf\tilde j}{\rm e}^{-i\omega t})$
the complex conductivity is defined by ${\bf\tilde j}=\sigma{\bf\tilde  E}$.
In the GR model, balancing forces as in \eqref{BoF} and using
the Josephson Relation \eqref{Josephson} to interpret the experimental
data, one concludes that $\nu \neq 0$ and there obtains an imaginary
part to the conductivity.  From the TDGL point of view, again using
\eqref{BoF} but with the Inertial Josephson Relation \eqref{inerJos},
one obtains a complex conductivity without recourse to postulating a
pinning potential.

While excellent experimental agreement has been attained through
more empirical methods such as the interpolation between normal
and superconducting states of Coffey and Clem \cite{CC91}, we would like to
stress that our derivation of the IJR relies only on
the same basic physical principles as the Josephson Relation.

Using ${\bf\tilde r}_{\rm L}=i{\bf\tilde v}_{\rm L}/\omega$, 
we substitute the vortex velocity ${\bf\tilde v}_{\rm L}$ from 
\eqref{BoF} into the Josephson Relation \eqref{Josephson}, which yields 
\begin{equation}
{\bf\tilde E}=\frac{\Phi_0 {\bf B}\times {\bf\tilde j} \times{\bf B} }{(\eta+i\nu/\omega) c^2B} 
\end{equation}
Since the magnetic field is perpendicular to the current, the 
double vector product becomes 
${\bf\tilde E}= \Phi_0 B{\bf\tilde j} / ((\eta+i\nu/\omega) c^2)$. 
The conductivity 
\begin{equation}
\sigma_{\rm JR}=\frac{(\eta+i\nu/\omega) c^2}{\Phi_0 B}
\end{equation}
has non-zero imaginary part exclusively due to the pinning; $\nu\neq0$. 
As one can see in \figref{fig_dataplot}, this model allows for a qualitatively 
good fit of experimental data since the imaginary part 
decreases as $1/\omega$ and increases with decreasing magnetic field.

Now let us turn to the application of the Inertial Josephson Relation \eqref{inerJos} 
and show that the inertial term leads to complex conductivity even in absence of pinning.
We evaluate the vortex velocity $\vl$ from the 
balance equation \eqref{BoF} with $\nu=0$ and substitute it into the Inertial Josephson Relation
\eqref{inerJos}. The resulting electric field is
${\bf\tilde E}=(\Phi_0 B/(\eta c^2)) {\bf\tilde j}-i\omega (m^*/( {e^*}^2 n)){\bf\tilde j}$ 
giving a conductivity of the Drude type
\begin{equation}
\label{drudecond}
\sigma_{\rm IJR}=\frac{\eta c^2}{ \Phi_0 B_{c2}}
\bigg( \frac{B}{B_{c2}} - i\omega\tau \bigg)^{-1}       
\end{equation}
where $B_{c2}$ is the upper critical field at zero
temperature and $\tau=\eta m^* c^2 /({e^*}^2 n\Phi_0 B_{c2})$
is the relaxation time of the condensate. 

In \figref{fig_dataplot} we compare the imaginary part of conductivity 
\eqref{drudecond} with experimental data of Ikebe {\em et al} \cite{ISIFK09} 
using the relaxation time $\tau=\pi^4\kappa^2\Phi_0\sigma_{\rm N}/
(14\zeta(3)c^2B_{c2})$ in the dirty limit \cite{Kop01}. 
The experimentally established normal conductivity is
$\sigma_{\rm N}=2\times 10^4 \ \Omega^{-1}{\rm cm}^{-1}$ and the upper critical field 
$B_{c2}=12.2$T, as in \cite{ISIFK09}.
                                
The GL parameter is $\kappa=38$, usual for NbN. 
From $\tau$ we find the friction coefficient
${\eta}=\pi^4\Phi_0 B_{c2} n \sigma_{\rm N}/(14\zeta(3)n_0 c^2)$. 
We have used the London penetration depth at 
zero temperature $\lambda^2=m^*c^2/(2\pi e^{*2}n_0)$, the GL 
coherence length $\xi=\lambda/\kappa$, and 
$B_{\rm c2}=\Phi_0/(2\pi\xi^2)$. 
The density of Cooper pairs we evaluate from the 
TDGL solution in the small signal limit; $ n / n_0 = ( 1 - T/T_c - B/B_{c2} ) / (2\beta_{\rm A})$
with the Abrikosov coefficient $\beta_{\rm A}=1.16$ for 
the triangular lattice. Using $\zeta(3)=1.202$ we obtain
$\sigma=2.495\,\sigma_{\rm N}( 1 - T/T_c - B/B_{c2} )/( B/B_{c2} - i\omega\tau)$.
A good agreement of the conductivity \eqref{drudecond} with
experimental data shows that the set of equations \eqref{BoF} and
\eqref{inerJos} captures the FIR response without any need for the 
addition of pinning to the theory.

Let us revisit Josephson's derivation of relation \eqref{Josephson} in
light of the present discussion. Josephson assumed a steady motion of
vortices and condensate. Since the condensate is not permanently
accelerated by passing vortices, he concluded that the electromotive
force is balanced by a gradient of the electrochemical
potential. Josephson's relation \eqref{Josephson} represents the
balance of these two forces. Away from low frequencies, it becomes
important that the condensate is continually accelerated by the
alternating field.  The balance of forces then leads to the Inertial
Josephson Relation \eqref{inerJos}.  Finally it is important to note,
as did Josephson, that the ${\bf E}'$ appearing in his relation is not
exactly the electric field ${\bf E}$ appearing in the present paper;
the subtle difference is that Josephson's ${\bf E}'$ depends on the
chemical potential rather than the electromagnetic scalar potential.

The derivation here is restricted to the case of an infinitesimally
thin layer.  It is reasonable to expect that its validity is more
general, although the above analysis benefits from the simplicity of
averaging over the uniform electric and magnetic field
configuration. We did not consider the conductivity contribution of
normal electrons; since the imaginary part of the normal conductivity
is negligible, this is not important for our discussion.

We have made use of Ginzburg-Landau theory to provide a careful
derivation of the Inertial Josephson Relation, which relates electric
and magnetic fields to not only vortex velocity, as in the Josephson
Relation, but also to the acceleration of the charge carriers.  We
have shown that the same reasoning which leads to the Josephson
Relation at low frequencies can be used to carefully derive the more
generally applicable inertial version, first suggested by Abrikosov
{\em et al} \cite{AKK65}.

We have illustrated the use of the Inertial Josephson Relation in
interpreting experimental data in the context of high-frequency
response. While the introduction of pinning is required to explain the
data if the usual Josephson Relation is used, we have argued that in
fact this use is erroneous; the Inertial Josephson Relation should be
utilised at high frequencies, and this then does not require
pinning. The analysis suggests that it is the inertial term which is
responsible for the imaginary part of the conductivity in the
high-frequency region.

\acknowledgments
This work was supported by research plan
MSM 0021620834, by grants
GA\v{C}R 204/10/0687 and 202/08/0326, and 
GAAV 100100712, by DAAD and by Taiwan-Czech PPP project: 99-2911-I-216-001.

\bibliography{biblio/tdgl,biblio/sem1,biblio/bose,biblio/delay2,biblio/delay3,biblio/gdr,biblio/genn,biblio/chaos,biblio/kmsr,biblio/kmsr1,biblio/kmsr2,biblio/kmsr3,biblio/kmsr4,biblio/kmsr5,biblio/kmsr6,biblio/kmsr7,biblio/micha,biblio/refer,biblio/sem2,biblio/sem3,biblio/short,biblio/spin,biblio/spin1,biblio/solid,biblio/deform}

\begin{thebibliography}{12}
\expandafter\ifx\csname natexlab\endcsname\relax\def\natexlab#1{#1}\fi
\expandafter\ifx\csname bibnamefont\endcsname\relax
  \def\bibnamefont#1{#1}\fi
\expandafter\ifx\csname bibfnamefont\endcsname\relax
  \def\bibfnamefont#1{#1}\fi
\expandafter\ifx\csname citenamefont\endcsname\relax
  \def\citenamefont#1{#1}\fi
\expandafter\ifx\csname url\endcsname\relax
  \def\url#1{\texttt{#1}}\fi
\expandafter\ifx\csname urlprefix\endcsname\relax\def\urlprefix{URL }\fi
\providecommand{\bibinfo}[2]{#2}
\providecommand{\eprint}[2][]{\url{#2}}

\bibitem[{\citenamefont{Blatter et~al.}(1994)\citenamefont{Blatter, Feigel'man,
  Geshkenbein, Larkin, and Vinokur}}]{BFGLV94}
\bibinfo{author}{\bibfnamefont{G.}~\bibnamefont{Blatter}},
  \bibinfo{author}{\bibfnamefont{M.~V.} \bibnamefont{Feigel'man}},
  \bibinfo{author}{\bibfnamefont{V.~B.} \bibnamefont{Geshkenbein}},
  \bibinfo{author}{\bibfnamefont{A.~I.} \bibnamefont{Larkin}},
  \bibnamefont{and} \bibinfo{author}{\bibfnamefont{V.~M.}
  \bibnamefont{Vinokur}}, \bibinfo{journal}{Rev. Mod. Phys.}
  \textbf{\bibinfo{volume}{66}}, \bibinfo{pages}{1125} (\bibinfo{year}{1994}).

\bibitem[{\citenamefont{Rosenstein and Li}(2010)}]{RL09}
\bibinfo{author}{\bibfnamefont{B.}~\bibnamefont{Rosenstein}} \bibnamefont{and}
  \bibinfo{author}{\bibfnamefont{D.}~\bibnamefont{Li}}, \bibinfo{journal}{Rev.
  Mod. Phys.} \textbf{\bibinfo{volume}{82}}, \bibinfo{pages}{109}
  (\bibinfo{year}{2010}).

\bibitem[{\citenamefont{Troyanovski et~al.}(1999)\citenamefont{Troyanovski,
  Aarts, and Kes}}]{TAK99}
\bibinfo{author}{\bibfnamefont{A.~M.} \bibnamefont{Troyanovski}},
  \bibinfo{author}{\bibfnamefont{J.}~\bibnamefont{Aarts}}, \bibnamefont{and}
  \bibinfo{author}{\bibfnamefont{P.~H.} \bibnamefont{Kes}},
  \bibinfo{journal}{Nature} \textbf{\bibinfo{volume}{339}},
  \bibinfo{pages}{665} (\bibinfo{year}{1999}).

\bibitem[{\citenamefont{Gittleman and Rosenblum}(1966)}]{GR66}
\bibinfo{author}{\bibfnamefont{J.~I.} \bibnamefont{Gittleman}}
  \bibnamefont{and}
  \bibinfo{author}{\bibfnamefont{B.}~\bibnamefont{Rosenblum}},
  \bibinfo{journal}{Phys. Rev. Lett.} \textbf{\bibinfo{volume}{16}},
  \bibinfo{pages}{734} (\bibinfo{year}{1966}).

\bibitem[{\citenamefont{Josephson}(1965)}]{J65}
\bibinfo{author}{\bibfnamefont{B.~D.} \bibnamefont{Josephson}},
  \bibinfo{journal}{Phys. Lett.} \textbf{\bibinfo{volume}{16}},
  \bibinfo{pages}{242} (\bibinfo{year}{1965}).

\bibitem[{\citenamefont{Maniv et~al.}(2009)\citenamefont{Maniv, Rosenstein,
  Shapiro, and Shapiro}}]{MRSS09}
\bibinfo{author}{\bibfnamefont{T.}~\bibnamefont{Maniv}},
  \bibinfo{author}{\bibfnamefont{B.}~\bibnamefont{Rosenstein}},
  \bibinfo{author}{\bibfnamefont{I.}~\bibnamefont{Shapiro}}, \bibnamefont{and}
  \bibinfo{author}{\bibfnamefont{B.~Y.} \bibnamefont{Shapiro}},
  \bibinfo{journal}{Phys. Rev. B} \textbf{\bibinfo{volume}{80}},
  \bibinfo{pages}{134512} (\bibinfo{year}{2009}).

\bibitem[{\citenamefont{Lin and Matlock}(2010)}]{LM10}
\bibinfo{author}{\bibfnamefont{F.~P.-J.} \bibnamefont{Lin}} \bibnamefont{and}
  \bibinfo{author}{\bibfnamefont{P.}~\bibnamefont{Matlock}},
  \bibinfo{journal}{Phys. Rev. B} \textbf{\bibinfo{volume}{82}},
  \bibinfo{pages}{024516} (\bibinfo{year}{2010}).

\bibitem[{\citenamefont{Hocquet et~al.}(1992)\citenamefont{Hocquet, Mathieu,
  and Simon}}]{HMS92}
\bibinfo{author}{\bibfnamefont{T.}~\bibnamefont{Hocquet}},
  \bibinfo{author}{\bibfnamefont{P.}~\bibnamefont{Mathieu}}, \bibnamefont{and}
  \bibinfo{author}{\bibfnamefont{Y.}~\bibnamefont{Simon}},
  \bibinfo{journal}{Phys. Rev. B} \textbf{\bibinfo{volume}{46}},
  \bibinfo{pages}{1061} (\bibinfo{year}{1992}).

\bibitem[{\citenamefont{Abrikosov et~al.}(1965)\citenamefont{Abrikosov,
  Kemoklidze, and Khalatnikov}}]{AKK65}
\bibinfo{author}{\bibfnamefont{A.~A.} \bibnamefont{Abrikosov}},
  \bibinfo{author}{\bibfnamefont{M.~P.} \bibnamefont{Kemoklidze}},
  \bibnamefont{and} \bibinfo{author}{\bibfnamefont{I.~M.}
  \bibnamefont{Khalatnikov}}, \bibinfo{journal}{Sov. Phys. JETP}
  \textbf{\bibinfo{volume}{21}}, \bibinfo{pages}{506} (\bibinfo{year}{1965}).

\bibitem[{\citenamefont{Bekarevich and Khalatnikov}(1961)}]{BK61}
\bibinfo{author}{\bibfnamefont{I.}~\bibnamefont{Bekarevich}} \bibnamefont{and}
  \bibinfo{author}{\bibfnamefont{I.~M.} \bibnamefont{Khalatnikov}},
  \bibinfo{journal}{Sov. Phys. JETP} \textbf{\bibinfo{volume}{13}},
  \bibinfo{pages}{643} (\bibinfo{year}{1961}).

\bibitem[{\citenamefont{Kopnin}(2001)}]{Kop01}
\bibinfo{author}{\bibfnamefont{N.~B.} \bibnamefont{Kopnin}},
  \emph{\bibinfo{title}{Theory of Nonequilibrium Superconductivity}}
  (\bibinfo{publisher}{Claredon Press}, \bibinfo{address}{Oxford},
  \bibinfo{year}{2001}).

\bibitem[{\citenamefont{Ikebe et~al.}(2009)\citenamefont{Ikebe, Shimano, Ikeda,
  Fukumura, and Kawasaki}}]{ISIFK09}
\bibinfo{author}{\bibfnamefont{Y.}~\bibnamefont{Ikebe}},
  \bibinfo{author}{\bibfnamefont{R.}~\bibnamefont{Shimano}},
  \bibinfo{author}{\bibfnamefont{M.}~\bibnamefont{Ikeda}},
  \bibinfo{author}{\bibfnamefont{T.}~\bibnamefont{Fukumura}}, \bibnamefont{and}
  \bibinfo{author}{\bibfnamefont{M.}~\bibnamefont{Kawasaki}},
  \bibinfo{journal}{Phys. Rev. B} \textbf{\bibinfo{volume}{79}},
  \bibinfo{pages}{174525} (\bibinfo{year}{2009}).

\bibitem[{\citenamefont{Coffey and Clem}(1991)}]{CC91}
\bibinfo{author}{\bibfnamefont{Mark W.}~\bibnamefont{Coffey}} \bibnamefont{and}
  \bibinfo{author}{\bibfnamefont{John R.}~\bibnamefont{Clem}}, \bibinfo{journal}
  {Phys. Rev. Lett.} \textbf{\bibinfo{volume}{67}}, \bibinfo{pages}{386}
  (\bibinfo{year}{1991}).

\end{thebibliography}

\end{document}